\documentclass[prb,twocolumn,showpacs,amssymb,amsmath]{revtex4}

\usepackage{graphicx}
\usepackage{bm}
\def\be{\begin{equation}}
\def\ee{\end{equation}}
\def\bea{\begin{eqnarray}}
\def\eea{\end{eqnarray}}
\newcommand{\matr}[4]{{\left(\begin{array}{cc} #1&#2\\#3&#4\\\end{array}\right)}}
\newcommand{\vect}[2]{{\left(\begin{array}{c} #1\\#2\\\end{array}\right)}}
\renewcommand{\vec}{\mathbf}

\renewcommand{\vr}{\vec{r}}

\newcommand{\vsigma}{\mbox{\boldmath $\sigma$}}

\newcommand{\vnabla}{\mbox{\boldmath $\nabla$}}

\begin{document}
\title{Resonant scattering in graphene with a gate-defined chaotic quantum dot}

\author{Martin Schneider}
\author{Piet W. Brouwer}
\affiliation{
Dahlem Center for Complex Quantum Systems and Institut f\"ur theoretische Physik,
Freie Universit\"at Berlin, Arnimallee 14, 14195 Berlin, Germany
}

\begin{abstract}
We investigate the conductance of an undoped graphene sheet with two metallic contacts and an electrostatically gated island (quantum dot) between the contacts. Our analysis is based on the Matrix Green Function formalism, which was recently adapted to graphene by Titov {\em et al.} [Phys.\ Rev.\ Lett.\ {\bf 104}, 076802 (2010)]. We find pronounced differences between the case of a stadium-shaped dot (which has chaotic classical dynamics) and a disk-shaped dot (which has integrable classical dynamics) in the limit that the dot size is small in comparison to the distance between the contacts. In particular, for the stadium-shaped dot the two-terminal conductance shows Fano resonances as a function of the gate voltage, which cross-over to Breit-Wigner resonances only in the limit of completely separated resonances, whereas for a disk-shaped dot sharp Breit-Wigner resonances resulting from higher angular momentum remain present throughout. 
\end{abstract}

\maketitle

\section{Introduction}
With its remarkable electronic properties, graphene, a single layer of carbon atoms as they occur in graphite, continues to be the subject of theoretical and experimental research.\cite{kn:castroneto2009,kn:geim2009,kn:dassarma2011} Graphene owes its special electronic properties to a Dirac-like dispersion in the absence of impurities, with the Fermi level at the Dirac point in the absence of external doping. Although the density of states vanishes at the Dirac point, transport in clean undoped graphene is possible through evanescent modes, giving rise to a finite conductivity $4 e^2/\pi h$ at zero temperature.\cite{kn:fradkin1986,kn:ludwig1994,kn:ziegler1998,kn:shon1998,kn:tworzydlo2006,kn:peres2006,kn:katsnelson2006} 
(At finite temperature, the conductivity of clean undoped graphene is formally divergent if effects of electron-electron interactions are neglected. \cite{kn:kashuba2008,kn:fritz2008,kn:mueller2009})

If the Fermi level $\varepsilon_{\rm F}$ is moved away from the Dirac point $\varepsilon=0$, {\em e.g.}, by application of a gate voltage, a finite concentration of charge carriers is created, for positive as well as negative $\varepsilon_{\rm F}$. It is for this reason that generically charge carriers in graphene can not be confined electrostatically, or that the addition of charged impurities to otherwise undoped graphene leads to an enhancement of the conductivity, rather than a suppression.\cite{kn:bardarson2007,kn:nomura2007,kn:sanjose2007,kn:adam2009} (For this effect to occur it is essential that the charged impurities are at a finite distance from the graphene sheet, so that they do not couple the Fermi points.\cite{kn:nomura2007b,kn:adam2007})

Recently, Bardarson, Titov and one of the authors showed that there is an exception to the rule that electrons in graphene can not be confined with the help of gate voltages, if the shape of the gate-defined region is such, that the Dirac equation is separable.\cite{kn:bardarson2009} An example of a gate-defined region (or ``quantum dot'') for which the Dirac equation is separable is a disk-shaped dot with a uniform potential inside the disk. In this case, electronic states have a well-defined orbital angular momentum. Electrons with the minimal angular momentum are not confined, whereas electrons with a higher angular momentum are bound to the quantum dot. The confinement is perfect if the graphene sheet surrounding the dot is undoped (with $\varepsilon_{\rm F} = 0$); otherwise the bound states become resonances.\cite{kn:matulis2008,kn:hewageegana2008} If the shape of the quantum dot is such, that the Dirac equation is not separable, as is the case for a stadium-shaped gated region, no bound states exist. 

In a semiclassical picture, the difference between integrable and non-integrable shapes can be understood with the help of Klein tunneling:\cite{kn:klein1929,kn:dombey1999} An electron incident on an electrostatic barrier in graphene is transmitted perfectly at perpendicular incidence, while the reflection probability sharply increases away from perpendicular incidence.\cite{kn:cheianov2006,kn:katsnelson2006,kn:beenakker2008} For a quantum dot with a shape, such that the classical dynamics is integrable, the angle of incidence is fixed, or only a limited set of angles appear in an electron's orbit, which allows for the confinement.\cite{kn:silvestrov2007,kn:shytov2008} On the other hand, with non-integrable classical dynamics, every electron eventually will approach the interface at perpendicular incidence and escape. (While this semiclassical argument explains the qualitative difference between integrable and non-integrable gate-defined quantum dots, it fails to account for details of the bound or quasi-bound states, because it does not take into account the Berry phase, which leads to half-integral angular momenta in graphene.\cite{kn:recher2007})

\begin{figure}
\includegraphics[width=2.5in]{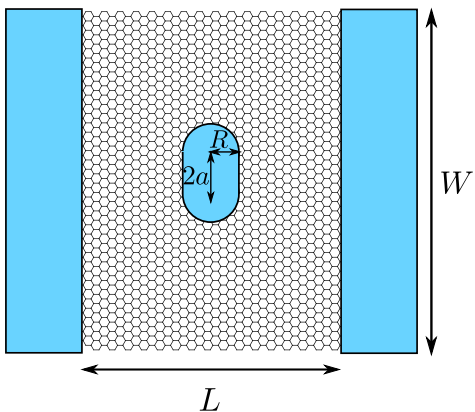}
\caption{(Color online) Stadium quantum dot (parameters $R$, $a$) in a rectangular graphene sample of dimensions $L\times W$. The sample is attached to metallic leads on the left and right side, defining a two-terminal geometry.}
\label{fig:stadium}
\end{figure}

As shown in Ref.\ \onlinecite{kn:bardarson2009}, the existence of bound states can be revealed through a two-terminal measurement of the conductance of an undoped graphene sheet with a gate-defined quantum dot (see Fig. \ref{fig:stadium}),  which shows narrow resonances for the bound states, and wider resonances for the quasibound states. The difference between integrable quantum dots (with bound states) and non-integrable quantum dots (without bound states) becomes most pronounced in the limit that the distance $L$ between the two metallic contacts becomes large. However, a detailed investigation of this limit was not possible in Ref.\ \onlinecite{kn:bardarson2009} because of limitations of the numerical approach required to study the non-integrable case.

Recently, the problem of gate-defined quantum dots in graphene was revisited by Titov, Ostrovsky, Gornyi, Schuessler, and Mirlin.\cite{kn:titov2010} These authors adapted the matrix Green function method, originally developed by Nazarov in the context of mesoscopic superconductivity,\cite{kn:nazarov1994} to two-terminal transport in graphene. One of their main results is a relation between the two-terminal conductance of a rectangular graphene sheet containing an arbitrary ``scatterer'' and the scatterer's $T$-matrix. A gate-defined quantum dot is a special case of such a scatterer. Using this method, Titov {\em et al.} were able to give an analytic expression for the conductance of an otherwise undoped graphene sheet with a disk-shaped quantum dot, reproducing the numerically obtained conductance of Bardarson {\em et al}.\cite{kn:bardarson2009} The usefulness of the matrix Green function method has also been demonstrated by its application to disordered graphene.\cite{kn:schuessler2010,kn:ostrovsky2010}

The purpose of the present article is to find out what information the matrix Green function method makes available if we consider undoped graphene with a non-separable quantum dot. This requires a numerical calculation of the $T$ matrix, since no analytical results are available for a quantum dot for which the Dirac equation is not separable. We here show that the numerical evaluation of the $T$ matrix can be done very efficiently and that the remaining steps in the matrix Green function formalism can still be carried out analytically. In particular, once the $T$ matrix of the quantum dot is known, the $L$ dependence of the conductance resonances can be found without additional numerical effort. This allows us to obtain information about the constitution of the resonances deep into the regime in which resonances are separated, which could not be addressed with the generic numerical algorithm used in Ref.\ \onlinecite{kn:bardarson2009}. We apply the formalism to a stadium-shaped quantum dot. Our main result is that the resonance lineshapes are described by the Fano resonance formula in the case of the stadium-shaped ({\em i.e.,} non-separable) quantum dot, while they are Breit-Wigner resonances for the disk-shaped dot.

This paper is structured as follows: In Section \ref{sec:Methods} we describe the matrix Green function method and its numerical implementation. We present our numerical results for a stadium-shaped dot in section \ref{sec:Results}. We conclude in section \ref{sec:Conclusion}. Some technical details are provided in the appendices.

\section{Model and method}
\label{sec:Methods}

Our setup consists of a rectangular sample of graphene of length $L$ and width $W \gg L$, described by the Dirac Hamiltonian
\begin{equation}
  H=-i\hbar v \vsigma \cdot \vnabla + V(x)+U(\mathbf{r}),
  \label{eq:Dirac}
\end{equation}
where $\vsigma=(\sigma_x,\sigma_y)$ are Pauli matrices, $v$ is the velocity of the Dirac quasiparticles, and $U$ and $V$ are gate potentials. The potential $V(x)$ accounts for the metallic leads and is set to be $V(x)\rightarrow -\infty$ for $x<0$ and $x>L$, whereas $V(x) = 0$ for $0 < x < L$, so that the graphene sheet is tuned to the Dirac point in the region of interest; The potential $U(\vr)$ defines the quantum dot. It takes a constant value $U(\vr) = U$ inside the quantum dot, and $U(\vr) = 0$ outside the dot. 

In order to calculate the two-terminal conductance of the system, we apply Nazarov's matrix Green function formalism,\cite{kn:nazarov1994} which has been adapted to graphene by Titov {\em et al.}\cite{kn:titov2010} Below, we summarize the essential elements of this method. Central object is a $4 \times 4$ matrix Green function $\check{G}$ with matrix structure in the pseudospin space [corresponding to the Pauli matrices $\sigma_x$ and $\sigma_y$ in the Dirac equation (\ref{eq:Dirac})] and the ``retarded-advanced'' (RA) space. The matrix Green function is defined by the equation 
\begin{equation}
 \left(
 \begin{array}{cc}
  \varepsilon-H+i0 &-\hbar v \sigma_x \zeta \delta(x)\\
  -\hbar v \sigma_x\zeta\delta(x-L) &  \varepsilon-H-i0\\
 \end{array}
 \right)
 \check{G}(\mathbf{r},\mathbf{r'})=\delta(\mathbf{r}-\mathbf{r'}),
  \label{eq:checkGdefinition}
\end{equation}
where $\varepsilon$ is the quasiparticle energy (set to zero in the remainder of the calculation), and $\zeta$ is an additional parameter. 
Following Ref.\ \onlinecite{kn:titov2010}, we define the generating function $\mathcal{F}(\phi)$ as
\begin{equation}
  \mathcal{F}(\phi)= \left. \mathrm{Tr}\ln\check{G}^{-1} \right|_{\zeta = \sin(\phi/2)},
\end{equation}
where the trace operation includes summation over RA and pseudospin indices, as well as integration over spatial coordinates. The generating function $\mathcal{F}$ contains information about the full counting statistics of two-terminal transport through the system.\cite{kn:nazarov1994} Here, we restrict our diskussion to the two-terminal conductance $G$, which is obtained from $\mathcal{F}$ through the equality
\begin{equation}
 G=- 2 g_0 \left. \frac{\partial^2\mathcal{F}}{\partial \phi^2}\right|_{\phi=0}, \label{eq:Gequation}
\end{equation}
where $g_0=4e^2/h$ is the conductance quantum in graphene.

Denoting the matrix Green function for the graphene sheet without quantum dot ($U=0$) as $\check{G}_0$, and writing $\mathcal{F}_0 = \mathrm{Tr}\ln\check{G}_0^{-1}$, we rewrite Eq.\ (\ref{eq:checkGdefinition}) as
\begin{eqnarray}
  \delta \mathcal{F} &=& \mathcal{F} - \mathcal{F}_0 \nonumber \\ &=&
  \mathrm{Tr}\ln(1-U\check{G}_0).
\end{eqnarray}
The matrix Green function $\check{G}_0$ for the clean system (without dot) has been calculated analytically in Ref. \onlinecite{kn:titov2010}. The explicit expression is rather lengthy and can be found in Appendix \ref{app:MGF}. Substituting the explicit expression for $\check{G}_0$ into Equation (\ref{eq:Gequation}) then gives
\begin{equation}
  G = \frac{g_0 W}{\pi L} + \delta G,\ \
  \delta G = -2 g_0 \left. \frac{\partial^2 \delta \mathcal{F}}{\partial \phi^2}\right|_{\phi=0},
\end{equation}
where the first term corresponds to the conductivity of clean graphene \cite{kn:ludwig1994,kn:ziegler1998,kn:peres2006,kn:tworzydlo2006} and the second term gives the correction from the presence of the quantum dot, which is the focus of our calculation.

The calculation of Ref.\ \onlinecite{kn:titov2010} proceeds by expressing $\delta \mathcal F$ in terms of the $T$ matrix of the quantum dot. Hereto, one introduces the Green function $g$ for zero-energy quasiparticles in an infinite sample ({\em i.e.}, with $V = U = 0$),
\begin{equation}
 g(\mathbf{r},\mathbf{r'})=-\frac{i}{2\pi\hbar v}\frac{\vsigma \cdot(\mathbf{r}-\mathbf{r'})}{|\mathbf{r}-\mathbf{r'}|^2},
\end{equation}
as well as the function 
\begin{equation}
  \check{G}_{\rm reg}=\check{G}_0-g,
\end{equation}
which, being a difference of two Green functions, is regular if the spatial arguments coincide. With the standard Born series for the $T$ matrix,\cite{kn:messiah1961} 
\begin{equation}
  T=(1-Ug)^{-1}U,
\end{equation}
one finds\cite{kn:titov2010}
\begin{equation}
 \label{eq:F}
 \delta\mathcal{F}=\mathrm{Tr}\ln(1-T\check{G}_{\rm reg}),
\end{equation}
up to terms that do not depend on the counting field $\phi$. Equation \eqref{eq:F} is the basis for our further investigations.

We will consider the limit that the size of the quantum dot is small in comparison to the length $L$. In that case, it is advantageous to expand $\check{G}_{\rm reg}$ around the position $\vr_0$ of the center of the quantum dot. After a short algebraic manipulation, which is carried out in Appendix \ref{app:MGF}, one finds that one may replace the operators $T$ and $\check{G}_{\rm reg}$ in Eq.\ (\ref{eq:F}) by matrices with elements $T^{\mu\nu}$ and $G_{\rm reg}^{\mu\nu}$, with $\mu,\nu = 0,1,2,\ldots$ and 
\begin{eqnarray}
  G_{\rm reg}^{\mu\nu}&=&
  \left.
  \frac{\partial^{\mu}}{\partial x^{\mu}}
  \frac{\partial^{\nu}}{\partial x'^{\nu}}
  \check{G}_{\rm reg}(\vec{r},\vec{r}')
  \right|_{\vr=\vr'=\vr_0},
 \label{eq:Gmunu} \\
 \label{eq:Tmunu}
 T^{\mu\nu} &=& \frac{1}{\mu! \nu!}
  \int d^2\vec{r} d^2\vec{r'} [(x-x_0)-i\sigma_z(y-y_0)]^{\mu} \nonumber \\ &&
  \times T(\vec{r},\vec{r'}) [(x'-x_0)+i\sigma_z(y'-y_0)]^{\nu}.
\end{eqnarray}
Note that each element $G_{\rm reg}^{\mu\nu}$ is a $4\times4$-matrix with non-trivial operation on pseudospin and RA degrees of freedom, whereas $T^{\mu\nu}$ acts in pseudospin space only and leaves the RA space untouched. With this replacement, the trace in Eq.\ (\ref{eq:F}) is taken over the matrix indices $\mu$ or $\nu$, the pseudospin degrees of freedom, and the RA degrees of freedom.

Essentially, the matrix $T^{\mu\nu}$ is the $T$ matrix in a partial-wave expansion. In graphene, the partial wave expansion involves waves of angular momentum $\sigma(\mu + 1/2)$, with $\sigma = \pm 1$ a pseudospin index and $\mu = 0,1,2,\ldots$ the index of the matrix $T^{\mu\nu}$. The corresponding basis functions $\psi_{k,\mu,\sigma}(\vr)$ are defined at a finite energy $\varepsilon = \hbar v k$ only, where, for definiteness, we choose $k > 0$. They are the solutions of the Dirac equation $H \psi_{k,\mu,\sigma}(\vr) = \varepsilon \psi_{k,\mu,\sigma}(\vr)$ in the absence of the potentials $U$ and $V$. With the matrix notation $\Psi_{k\mu} = (\psi_{k,\mu,+1},\psi_{k,\mu,-1})$, one has
\begin{widetext}
\begin{eqnarray}
  \Psi_{k\mu}(\vr) &=&
  \sqrt{\frac{k}{4 \pi}}
  \left( \begin{array}{cc} e^{i \mu \theta} J_{\mu}(k r) & i e^{-i (\mu+1)\theta} J_{\mu+1}(k r) \\
  i e^{i (\mu+1) \theta} J_{\mu+1}(k r) & e^{-i \mu \theta} J_{\mu}(k r) \end{array} \right), ~~~
  \label{eq:basis}
\end{eqnarray}
where $J_{\nu}$ is a Bessel function and we used polar coordinates $(r,\theta)$.
Denoting the $T$ matrix in the partial-wave basis by $T_{\mu \nu}(k)$, one then finds that
\begin{equation}
  T^{\mu\nu} = \lim_{k\rightarrow0}\frac{2^{\mu+\nu+2}\pi}{k^{\mu+\nu+1}}
  T_{\mu\nu}(k),
  \label{eq:Tmunumunu}
\end{equation}
where we again refer to appendix \ref{app:MGF} for details. In our discussion below, we will refer to the mode indices $\mu,\nu=0,1,2,\ldots$ as $s$, $p$, $d$, \ldots.

It remains to describe a method to calculate the $T$-matrix $T_{\mu\nu}(k)$ for a specific quantum dot potential $U(\vr)$. This is done through by first relating $T$ to the scattering matrix $S_{\mu\nu}(k)$ in the partial-wave basis,
\begin{equation}
  S_{\mu\nu}(k)=\delta_{\mu\nu}-2\pi i T_{\mu\nu}(k).
\end{equation}
The scattering matrix $S_{\mu\nu}(k)$ relates the coefficients of incoming ($-$) and outgoing ($+$) parts of the basis functions (\ref{eq:basis}) at energy $\varepsilon = \hbar v k$. Again using the shorthand notation $\Psi_{k\mu}^{(\pm)} = (\psi_{k,\mu,+1}^{(\pm)},\psi_{k,\mu,-1}^{(\pm)})$, these are 
\begin{eqnarray}
  \Psi_{k\mu}^{(\pm)}(\vr) &=&
  \sqrt{\frac{k}{4 \pi}}
  \left( \begin{array}{cc} 
  e^{i \mu \theta} H_{\mu}^{(\pm)}(k r) &
  i e^{-i (\mu+1) \theta} H_{\mu+1}^{(\pm)}(k r) 
  \\
  i e^{i (\mu+1) \theta} H_{\mu+1}^{(\pm)}(k r) &
  e^{-i \mu  \theta} H_{\mu}^{(\pm)}(k r)
  \end{array} \right),~~~
  \label{eq:basisinout}
\end{eqnarray}
\end{widetext}
where the $H_{\mu}^{(\pm)} = J_{\mu} \pm i Y_{\mu}$ are Hankel functions of the first ($+$) and second kind ($-$), respectively. The scattering matrix $S_{\mu\nu}(k)$ is then defined through the asymptotic form of the solution of the Dirac equation $H \psi_k(\vr) =\hbar v k \psi_k(\vr)$ for $r \to \infty$, which takes the general form
\begin{equation}
 \label{eq:wavesphere}
  \psi_k(\vr) = \sum_{\mu} \left(
  \Psi_{k\mu}^{(+)}(\vr) a_{k\mu} +
  \Psi_{k\mu}^{(-)}(\vr) b_{k\mu} \right),
\end{equation} 
with, in $2 \times 2$ matrix notation,
\begin{equation}
  a_{k\mu} = \sum_{\nu} S_{\mu\nu}(k) b_{k\nu}.
\end{equation}

To find $S_{\mu\nu}(k)$, we employ a variation of a method commonly applied to rectangular strips of disordered graphene.\cite{kn:bardarson2007} Hereto, we divide the graphene sheet in circular slices $a_{j-1} < r < a_{j}$, $j=1,\ldots,N$, with $a_{0} = 0$ and $a_{N}$ so large that $U(\vr) = 0$ for $r > a_{N}$. We then obtain the scattering matrix $S(k)$ by solving two auxiliary scattering problems first.

In the first auxiliary problem, we set the potential $U$ to zero everywhere, except in the circular slice $a_{j-1} < r < a_{j}$. In this case, the wavefunctions can be expanded in incoming and outgoing partial waves for $r < a_{j-1}$ as well as for $r > a_{j}$. The general solution of the Dirac equation in the regions $r < a_{j-1}$ and $r > a_{j}$ can be characterized by means of two reflection matrices $\rho_j$ and $\rho'_j$ and two transmission matrices $\tau_j$ and $\tau'_j$, such that $\rho$ relates the coefficients of outgoing partial waves on the exterior to the coefficients of incoming partial waves on the exterior etc. If $\delta a = a_{j}-a_{j-1}$ is sufficiently small, these matrices can be calculated in the first-order Born approximation. Writing $\rho_{\mu \nu} = \delta \rho_{\mu \nu}$, $\rho'_{\mu\nu} = \delta \rho'_{\mu\nu}$, $\tau_{\mu\nu} = \delta_{\mu\nu} + \delta \tau_{\mu\nu}$, and $\tau_{\mu\nu}' = \delta_{\mu\nu} + \delta \tau'_{\mu\nu}$, this calculation gives
\begin{eqnarray*}
  \delta \rho_{\mu\nu}
  &=& - \frac{i \pi}{2 \hbar v} \int_{a_{j-1} < r < a_{j}} d\vr\,
  \Psi_{k\mu}^{(+)}(\vr)^{\dagger}
  U(\vr) \Psi_{k\nu}^{(-)}(\vr), \\
  \delta \tau_{\mu\nu}
  &=& - \frac{i \pi}{2 \hbar v} \int_{a_{j-1} < r < a_{j}} d\vr\,
  \Psi_{k\mu}^{(-)}(\vr)^{\dagger}
  U(\vr) \Psi_{k\nu}^{(-)}(\vr), \\
  \delta \rho'_{\mu\nu}
  &=& - \frac{i \pi}{2 \hbar v} \int_{a_{j-1} < r < a_{j}} d\vr\,
  \Psi_{k\mu}^{(-)}(\vr)^{\dagger}
  U(\vr) \Psi_{k\nu}^{(+)}(\vr), \\
  \delta \tau'_{\mu\nu}
  &=& - \frac{i \pi}{2 \hbar v} \int_{a_{j-1} < r < a_{j}} d\vr\,
  \Psi_{k\mu}^{(+)}(\vr)^{\dagger}
  U(\vr) \Psi_{k\nu}^{(+)}(\vr), 
\end{eqnarray*}
up to corrections of higher order in $\delta a$.
  
In the second auxiliary problem, we set $U$ to zero for $r > a_{j}$ only. Defining $S_j(k)$ to be the scattering matrix for this situation, we obtain the recursion relation
\begin{equation}
  S_{j}(k) = \rho_{j} + \tau'_j (1 - S_{j-1}(k)  \rho'_{j})^{-1} S_{j-1}(k)\tau_{j}.
  \label{eq:Sjrecursion}
\end{equation}
Together with the initial condition $S_0 = 1$ and the equality $S_{N} = S$ this leads to the desired solution.

Unitarity implies the relations $\delta \rho'_j = - \delta \rho_j^{\dagger}$, $\delta \tau_j = -\delta \tau^{\dagger}_j$, and $\delta \tau'_j = - \delta \tau_j'^{\dagger}$. Consistent with the Born approximation, we may rewrite the recursion relation (\ref{eq:Sjrecursion}) as
\begin{equation}
  S_{j}(k) = S_{j-1}(k) [1 + i \delta h_j], \label{eq:smatmul}
\end{equation}
where
\begin{eqnarray}
  i \delta h_j &=& S^{\dagger}_{j-1}(k) \delta \rho_j 
  + \delta \rho'_j S_{j-1}(k)
  \nonumber \\ && \mbox{}
   + S^{\dagger}_{j-1}(k) \delta \tau'_j S_{j-1}(k) + \delta \tau_j
\end{eqnarray}
is a hermitian matrix.

Although this concludes our formal description of the method, there are a few issues regarding the numerical implementation that we also need to discuss:

(i) --- For the final result, we have to take the low-$k$ limit of the scattering matrix $S(k)$, which, at first sight, may be problematic because the Hankel functions appearing in the Born approximation for the matrices $\delta \rho$, $\delta \rho'$, $\delta \tau$, and $\delta \tau'$ diverge in this limit. However, this problem can easily be circumvented by shifting the potential $U(\vr) \to U(\vr) + \hbar v k'$ for $r < a_{N}$, while at the same time increasing the quasiparticle energy according to $k \to k + k'$. This means that we transfer contributions between the free propagation and the perturbation. We then solve the scattering problem inside the disk starting from ``free'' Dirac fermions with energy $\hbar v(k+k')$, which remains finite in the limit $k \rightarrow 0$. The wavefunction outside the disk, where free electrons have momentum $k$, can be found by matching to the wave function inside the disk, and contains the information of the scattering matrix we need.

For the low-$k$-expansion of the scattering matrix, we then find an expression of the form
\begin{equation}
 \label{eq:r-exp}
  S_{\mu\nu}(k)= \delta_{\mu\nu}
  +
  S'_{\mu\nu}(k') k^{\mu + \nu + 1}+\mathcal{O}(k^{\mu + \nu+2}),
\end{equation}
where $S'(k')$ depends on $a_N$, $k'$, and the scattering matrix $S_{N}(k')$ obtained for the modified problem where we have replaced $k$ by $k'+k$ and safely can take the limit $k \to 0$. A detailed derivation is given in Appendix \ref{sec:Transfer}. Note, that the expansion in Eq. \eqref{eq:r-exp} gives us precisely that order in $k$, that we need in order to obtain the matrix $T^{\mu\nu}$ in Eq.\ (\ref{eq:Tmunumunu}). 

The wavenumber $k'$ of the quasiparticles can be chosen arbitrarily, and we have verified that the result of our calculations do not depend on the choice of $k'$. However, one may exploit this freedom for a wise choose of $k'$. It should be not too small, because otherwise the Hankel functions in the spherical incoming and outgoing waves become too large, indicating the fact that particles are repelled from the origin and therefore the Born approximation looses its applicability. On the other hand, if the wave vector $k'$ is chosen too large, the effective potential $U + \hbar k' v$ gets large, too, and the thickness $\delta a$ of the slices has to be decreased. 

(ii) --- It is sufficient to choose the radius $a_0$ of the first slice such, that $U(\vr)$ is uniform for $r < a_0$. For this situation an exact solution is available. 

(iii) --- In order to guarantee numerical stability during multiplication of scattering matrices, we change Eq. \eqref{eq:smatmul} to
\begin{equation}
  S_j(k)=S_{j-1}(k) \left(1+i\frac{\delta h_j}{2}\right)\left(1-i\frac{\delta h_j}{2}\right)^{-1},
\end{equation}
which is valid up to corrections beyond the accuracy of Born approximation, but ensures that the scattering matrices remain unitary at all times.

(iv) --- In the calculation of the scattering matrix, we keep a total of $M$ modes. After the calculation of the full scattering matrix $S$, we then further truncate the scattering matrix upon calculation of the determinant in Eq. \eqref{eq:F}, keeping matrix elements $S^{\mu\nu}$ with $\mu,\nu=0,1,...,M'$ only. Keeping a large number of modes $M$ in the calculation of the scattering matrix is important in order to properly resolve the dynamics inside the quantum dot. The number of modes $M'$ required for the calculation of the conductance depends on the ratio of the size $L$ of the graphene sheet and the size of the quantum dot, and can be kept small if this ratio is large. The numerical calculation of the conductance has converged, when the result does no longer change upon increasing $M$, $M'$ or $N$. For the results of the conductance shown here, we kept channels up to $d$-wave ($M'=2$) only, which we found to be sufficient for the parameters chosen in our calculation.

\section{Two-terminal conductance with stadium-shaped quantum dot}
\label{sec:Results}

\subsection{Two-terminal conductance}

As a nontrivial application of the formalism laid out in the previous Section, we have calculated the two-terminal conductance of a stadium-shaped quantum dot in an otherwise undoped graphene steet. The quantum dot is placed halfway between the contacts. It is characterized by the radius $R$ of the circular pieces and the length $2a$ of the linear segment, see Fig.\ \ref{fig:stadium}. The gate voltage $U(\vr) = U$ for $\vr$ inside the quantum dot, and $U(\vr) = 0$ otherwise. In our calculations, we have included contributions up to $d$-wave in the $T$ matrix, whereas as many modes were used for the calculation of the scattering matrix $S$ as were necessary to reach convergence. (See the discussion at the end of the previous section.)

\begin{figure}[t]
\includegraphics[width=2.9in]{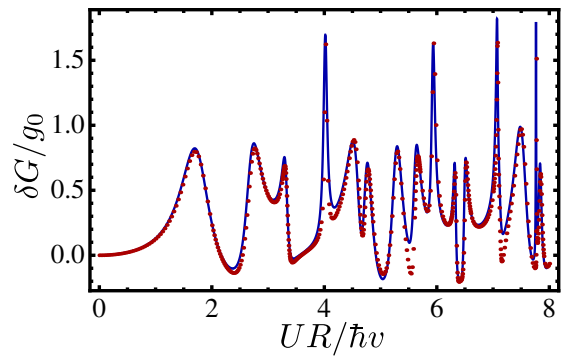}
\caption{(Color online) Conductance versus gate voltage $U$ for an undoped graphene sheet with a gate-defined stadium-shaped quantum dot. The parameters of the numerical calculation are $2a/R=\sqrt{3}$ and $L/R=5$. The curves correspond to the numerical calculation using the method of Sec.\ \ref{sec:Methods}, with contributions to the $T$ matrix up to $d$-wave (blue, solid), and to a direct numerical solution of the Dirac equation (red, points; data taken from Ref.\ \onlinecite{kn:bardarson2009}).}
\label{fig:resonances}
\end{figure}

In Fig.\ \ref{fig:resonances}, we show the two-terminal conductance as a function of the gate voltage $U$. The conductance exhibits several resonances due to the formation of quasi-bound states in the stadium. The parameters for the data in this figure are chosen to be the same as in Ref.\ \onlinecite{kn:bardarson2009}, so that the results can be compared. The excellent agreement between both curves concerns the position as well as the shape of the resonances. We attribute small deviations in the resonance heights to corrections from higher angular momentum that were not taken into account in the calculations. Such corrections disappear upon increasing the ratio of system size vs. dot size $L/R$ (see discussion below).

\begin{figure}[t]
\includegraphics[width=2.9in]{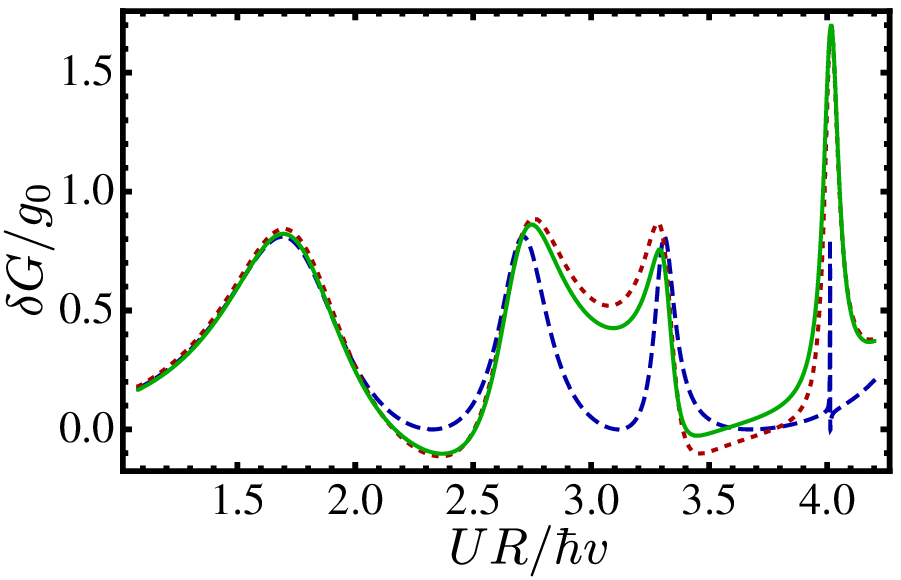}
\includegraphics[width=2.9in]{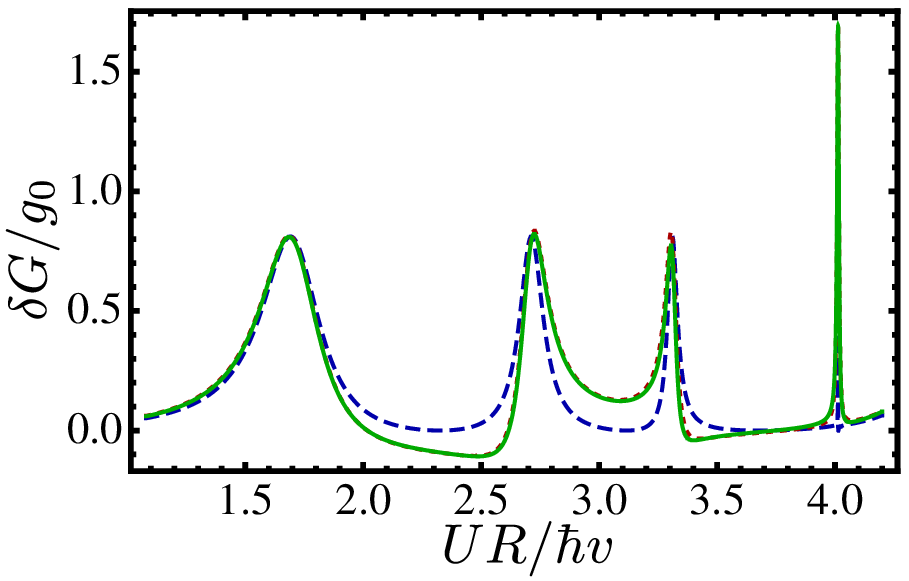}
\caption{(Color online) First four conductance resonances: The conductance has been calculated using Eq.\ (\ref{eq:F}) with the $T$ matrix truncated after the $s$-wave (blue, dashed), $p$-wave (red, dotted), and $d$-wave angular-momentum channel (green, solid), corresponding to $M'=0,1,2$, respectively. In the upper panel, we set $L/R=5$, while in the lower panel, $L/R=10$. The ratio $a/R$ remains the same as in Fig.\ \ref{fig:resonances}.}
\label{fig:resspd}
\end{figure}

The effect of inclusion of successive angular momentum channels in the $T$ matrix (while keeping essentially all angular momentum channels in the calculation of the scattering matrix $S$) is illustrated in Fig.\ \ref{fig:resspd}, where, in the upper panel, we show a close-up of the data of Fig.\ \ref{fig:resonances} with one, two, and three angular momentum channels included in the final step of the calculation (i.e. with truncation at $M'=0,1,2$). This figure clearly demonstrates that each resonance has contributions from more than one angular mode, as is to be expected for a chaotic quantum dot. In particular, we find that all resonances have a non-vanishing $s$-wave contribution, so that the resonance position can be extracted from the $s$-wave channel solely. For the data shown in Fig.\ \ref{fig:resonances}, one needs to include the $p$-wave contribution in order to obtain the correct resonance shape. The effect of the $d$-wave contribution is small, which we find remarkable, because the ratio of dot size $2 (R+a)$ versus contact size $L$ is $\approx 0.75$, which is not small in comparison to unity. The lower panel of Fig.\ \ref{fig:resspd} shows conductance data for $M'=0,1,2$, showing that the convergence with respect to $M'$ quickly improves upon increasing $L/R$.

We conclude, that most information about the resonances that is relevant for transport at large or moderately large $L/R$ is encoded in the contributions from small angular momentum, and is therefore stored in a small set of parameters. That fact is a strong indication of the power of the method employed here, in comparison to a direct simulation of the Dirac equation.

\subsection{Lineshape}

The main advantage of the present method over the direct numerical solution of the Dirac equation is that it allows one to extend the conductance calculations to the regime $L \gg R$: The complete dependence on the length $L$ is encoded on the Green function $\check{G}_{\rm reg}$, which is known analytically. It is only in the regime $L \gg R$ that resonances are well separated and can be characterized individually. The calculations of Ref.\ \onlinecite{kn:bardarson2009}, on the other hand, were limited to the regime $L \le 5 R$, where resonances were still strongly overlapping. (Reference \onlinecite{kn:bardarson2009} also considered the case $L \approx 8 R$ for $W=L$. However, in that case the conductance is strongly influenced by the finite width $W$ of the graphene sheet.)

\begin{figure}[t]
\includegraphics[width=2.9in]{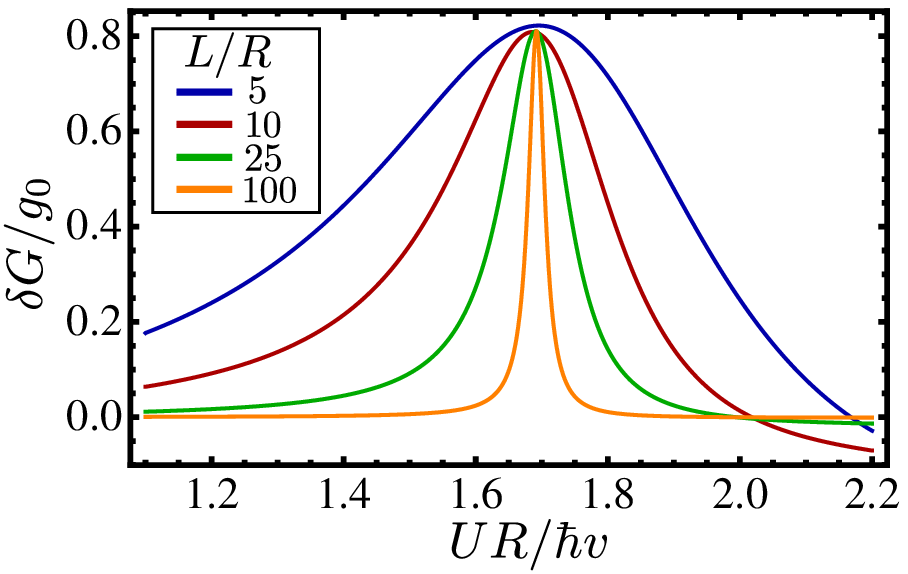}
\includegraphics[width=2.9in]{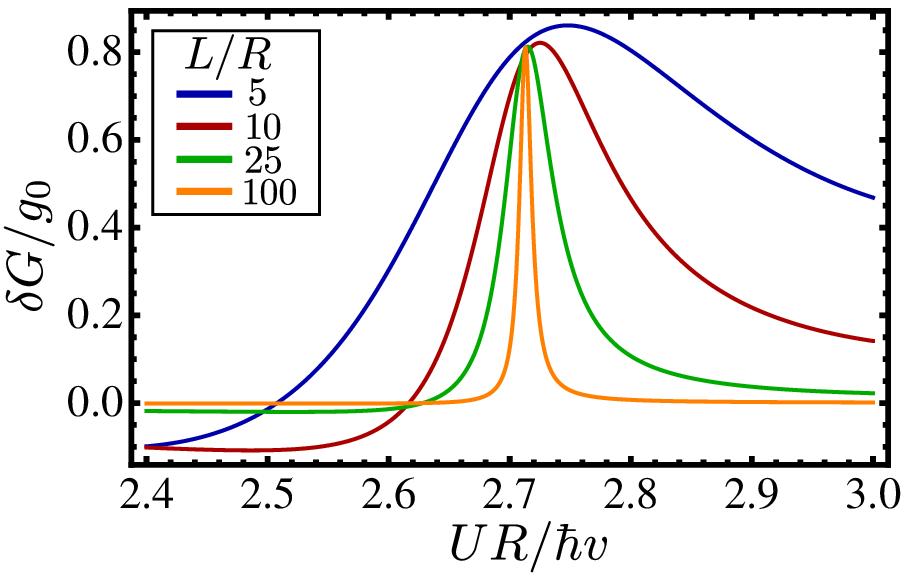}
\caption{(Color online) Dependence of the first and second resonance on the ratio of dot size versus system size $R/L$.}
\label{fig:resonancesRL}
\end{figure}

In Fig. \ref{fig:resonancesRL} we show how the first and second resonance behaves upon increasing the contact size $L$ at fixed dot size $R$. The plots illustrate that the resonance width shrinks, while the height saturates. The asymmetry of the line shape disappears in the limit $L/R \to \infty$, consistent with the expectation that $s$-wave scattering dominates if $L \gg R$. Similar behaviour is found for the other resonances. 

In order to quantitatively analyze the lineshapes, we note that each resonance is characterized by a divergence of the $T$ matrix. With a suitable parameterization of these divergencies an explicit expression for the resonance lineshape can be obtained. Hereto we introduce the dimensionless variable $\epsilon= U R/\hbar v$. Then, close to a divergence at $\tilde{\epsilon}_0$, the divergent part of $T^{\mu\nu}$ is of the form
\begin{equation}
 \label{eq:Tapp}
 T^{\mu\nu}(\epsilon)\simeq \frac{R^{\mu\nu}}{\epsilon-\tilde{\epsilon}_0},
\end{equation}
where the matrix $R^{\mu\nu}$ contains the information about the resonance shape. The matrix elements $R^{\mu\nu}$ are related by hermiticity, $R^{\mu\nu} = (R^{\nu\mu})^{\dagger}$ and time-reversal symmetry, $R^{\mu\nu}= \sigma_{y} (R^{\nu\mu})^{T} \sigma_y$. Moreover, for the specific problem we consider here, inversion symmetry and reflection symmetry impose further constraints on $R$, which allow us to parameterize the matrix as
\begin{equation}
  R = \left( \begin{array}{ccc}
  l_s \sigma_0 & i l_{sp} \sigma_x & l_{sd} \sigma_0 \\
  -i l_{sp} \sigma_x & l_p \sigma_0 & i l_{pd} \sigma_x \\
  l_{sd} \sigma_0 & -i l_{pd} \sigma_x & l_{d} \sigma_0 
  \end{array} \right),
  \label{eq:Rparam}
\end{equation}
where $\sigma_0$ is the $2 \times 2$ unit matrix.
Due to Kramers degeneracy, $R$ can be decomposed into two identical submatrices, which, for a generic resonance, are of unit rank. (In a chaotic quantum dot level repulsion ensures that this condition is always fulfilled.) This gives the additional constraints
\begin{equation}
  l_{sp}^2 = {l_s l_p},\ \
  l_{sd}^2 = {l_s l_d},\ \
  l_{pd}^2 = {l_p l_d}
  \label{eq:rankone}
\end{equation}
on the parameters in Eq.\ (\ref{eq:Rparam}). Note, that $l_{s}$ , $l_p$, and $l_d$ have the dimension of a length, length$^3$, and length$^5$, respectively.
 
Substituting Eq.\ \eqref{eq:Rparam} into Eq.\ \eqref{eq:F} and making use of the condition (\ref{eq:rankone}), we find that the line shapes are described by the Fano resonance formula \cite{kn:fano1961}
\begin{equation}
 \label{eq:Fano}
 G=G_{\rm nr}\frac{\left|2 (\epsilon-\epsilon_0)+q \Gamma\right|^2}{4(\epsilon-\epsilon_0)^2+\Gamma^2}
\end{equation}
Here, $G_{\rm nr}$ is the non-resonant conductance, $\Gamma$ is the resonance width, $\epsilon_0$ is the resonance position, and $q$ is the complex ``Fano parameter''. After subtracting the background conductance $G_{\rm nr}$, we rewrite Eq. \eqref{eq:Fano} as
\begin{equation}
 \delta G = g_0 \frac{\beta +4\alpha(\epsilon-\epsilon_0)}{4 (\epsilon-\epsilon_0)^2+ \Gamma^2},
\end{equation}
where $\beta = (G_{\rm nr}/g_0)(|q|^2-1) \Gamma^2$ and $\alpha = (G_{\rm nr}/g_0) \mbox{Re}\,q$.
For $\alpha=0$ the resonance has a Breit-Wigner shape; Non-zero $\alpha$ is responsible for an asymmetry in the lineshape. When we express the resonance parameters through the entries of $R$, we also use $\sigma_{sp}=\mathrm{sign}(l_{sp})$ etc., which is not fixed by Eq.\ (\ref{eq:rankone}):
\begin{eqnarray}
 \epsilon_0&=&\tilde{\epsilon}_0+\frac{\pi \sqrt{l_s l_p}\sigma_{sp}}{24 L^2}-\frac{7\pi^3 \sqrt{l_p l_d}\sigma_{pd}}{960 L^4},\\
 \Gamma&=&\left|\frac{l_{s}}{2 L}+\frac{\pi^2 l_p}{8 L^3}+\frac{\pi^2 \sqrt{l_s l_d}\sigma_{sd}}{4 L^3}+\frac{5\pi^4 l_d}{32 L^5}\right|,\\
 \beta
  &=&\frac{2 l_s^2}{\pi ^2L^2}+\frac{2 l_p l_s}{3 L^4}+\frac{8 \sqrt{l_s^3 l_d}\sigma_{sd}}{3 L^4}+\frac{19 l_p^2 \pi ^2}{72 L^6}\nonumber,\\
    & &+\frac{481 l_d l_s \pi ^2}{180 L^6}+\frac{17 l_p \sqrt{l_s l_d}\sigma_{sd} \pi ^2}{18 L^6}+\frac{61 l_d l_p \pi ^4}{90 L^8}\nonumber,\\
& &+\frac{251 \sqrt{l_s l_d^3}\sigma_{sd} \pi ^4}{180 L^8}+\frac{13549 l_d^2 \pi ^6}{28800 L^{10}}\\
  \alpha &=&-\frac{2 \sqrt{l_s l_p}\sigma_{sp}}{\pi L^2} + \frac{\pi \sqrt{l_p l_d}\sigma_{pd}}{2 L^4}.
\end{eqnarray}
Note, that the shift $\epsilon_0-\tilde \epsilon_0$ of the resonance position, as well as the asymmetry $\alpha$ is absent, if the resonance consists of a single angular momentum component only.

Now consider the situation, where we fix the product $UR$ of the gate voltage quantum dot and the quantum dot size --- {\em i.e.}, we look at a fixed resonance --- and increase the length $L$ of the graphene sheet. Then, as long as $l_{s}\neq 0$, at sufficiently large $L$, the conductance will be determined by its $s$-wave contribution, so that the resonance line shape reduces to the Breit-Wigner form. In this limit, the height for $\delta G$ approaches the constant universal value $8 g_0/\pi^2$, and the width scales as $\Gamma=l_{s}/{2L}$, as was found previously by Titov {\em et al.}\cite{kn:titov2010}

\subsection{Comparison Disc\,-\,Stadium}

We now summarize the main qualitative difference in the conductance resonances for stadium-shaped quantum dots, which were considered numerically here and in Ref.\ \onlinecite{kn:bardarson2009}, and disk-shaped dots, which were considered analytically in Ref.\ \onlinecite{kn:titov2010} and numerically in Ref.\ \onlinecite{kn:bardarson2009}. The main difference, which is illustrated in Fig.\ \ref{fig:resspd}, is that generic resonances for the stadium dot have contributions from more than one angular momentum channel. This means that the resonance lineshape changes from a Fano shape for moderate ratios $L/R$ to a Breit-Wigner lineshape for large $L/R$, as seen in Fig.\ \ref{fig:resonancesRL}. In the limit $L/R \to \infty$ of isolated resonances, the height of the conductance resonances takes the universal value $8 g_0/\pi^2$, whereas the width of all resonances scales proportional to $R/L$. 
Though we considered a stadium-shaped dot only, we believe, that the features found here are generic and shared by all chaotic dots.

This is to be contrasted to the situation of a circular quantum dot, where different angular momentum channels do not mix, and the resonances are of pure type ($s$-wave, $p$-wave,\ldots). The resonance lineshape is always of Breit-Wigner form and approaches a constant height upon taking the limit $L/R \to \infty$. The asymptotic resonance height depends on the angular momentum channel, with a height $8 g_0/\pi^2$ for $s$-wave resonances ($l = 0$) and $\lesssim 2 g_0$ for higher angular-momentum channels ($l \ge 1$). The resonance width depends on the angular momentum channel, $\Gamma \propto (R/L)^{2l+1}$. 

We close this comparison with a comment on the results of Ref.\ \onlinecite{kn:bardarson2009}. There it was found that the resonance height of the chaotic dot goes to zero upon taking the limit of large $L/R$. This observation referred to a setup with aspect ratio $W/L=1$ and periodic boundary conditions along the transverse direction. For aspect ratio $W\ll L$, only the mode with zero transverse momentum substantially contributes to transport.\cite{kn:tworzydlo2006} Therefore, $s$-wave scattering does not affect the conductance, resulting the observed suppressed height of $s$-wave conductance resonances. We verified that the same phenomenon occurs in our calculations, by evaluating the regularized Green function $\check{G}_{\rm reg}$ for finite width. This explains, why the resonances in Ref. \onlinecite{kn:bardarson2009} completely disappear in the limit of large $L/R$, while in our investigation for aspect ratio $W\gg L$ the height of the resonances remains finite.

\section{Conclusion}
\label{sec:Conclusion}
In this article, we investigated the resonances of the conductance of a graphene sheet with a chaotic quantum dot. Using a numerical implementation of the matrix Green function method of Ref.\ \onlinecite{kn:titov2010}, we were able to study the behaviour of the resonances in the limit of well-separated resonances. This essential limit could not be reached in the original treatment of the problem.\cite{kn:bardarson2009}

As was proposed in Ref.\ \onlinecite{kn:bardarson2009}, the resonances of the chaotic dot behave significantly different compared to the case when the gated region is circular. While the circular geometry does not allow for mode mixing, so that all resonances are Breit-Wigner resonances with a well-defined angular momentum, in the stadium dot, due to its non-integrable dynamics, all resonances have contributions from all scattering channels. The presence of mixed angular-momentum modes is responsible for an asymmetry of the line shape, described by the Fano resonance formula. In the limit of very well separated resonances, corresponding to the limit in which the size $R$ of the quantum dot is much smaller than the distance $L$ between the metallic contacts, for the chaotic quantum dot all resonances are dominated by the lowest ($s$-wave) angular momentum component, and recover the Breit-Wigner form. In contrast, for the disk-shaped dot, parametrically narrower resonances for higher angular momentum channels ($p$-wave, $d$-wave, etc.) persist in the limit of large $L/R$. This establishes clear signature that distinguishes regular and chaotic dynamics of gate-defined quantum dots from the conductance resonances.

We gratefully acknowledge discussions with J.\ H.\ Bardarson and M.\ Titov. This work is supported by the Alexander von Humboldt Foundation in the framework of the Alexander von Humboldt Professorship, endowed by the Federal Ministry of Education and Research (PWB) and by the German Research Foundation (DFG) in the framework of the Priority Program 1459 ``Graphene''.

\appendix

\begin{widetext}
\section{Matrix Green Function}
\label{app:MGF}

In this appendix, we give explicit expressions for some of the matrix Green function appearing in Sec.\ \ref{sec:Methods}. We follow the supplementary material of Ref.\ \onlinecite{kn:titov2010}, but our results for the $d$-wave channel go beyond that reference.

Using Pauli matrices $\tau_x$, $\tau_y$, and $\tau_z$ for the retarded-advanced (RA) degree of freedom, and with $\tau_0$ for the $2 \times 2$ unit matrix in the RA grading, the explicit expression for the matrix Green function $\check{G}_0$ is
\begin{eqnarray}
  \check{G}_0(x,x';y)=\frac{1}{4\hbar vL}\check{V}(x)\check{\Lambda}
  \check{\Lambda}_{\tau}
  \check{\Lambda}_{\sigma}
  \check{\Lambda}\check{V}^{-1}(x')
\end{eqnarray}
with $\check{\Lambda} = \matr{1}{0}{0}{0} \otimes \sigma_z+\matr{0}{0}{0}{1} \otimes \sigma_0$ and
\begin{eqnarray}
  \check{\Lambda}_{\tau} &=&
  \matr{i \cosh({\phi y}/{2 L})}{\sinh ({\phi y}/{2L})}{\sinh ({\phi y}/{2L})}{-i \cosh({\phi y}/{2 L})} \otimes \sigma_0, \\
  \check{\Lambda}_{\sigma} &=& \tau_0 \otimes
   \matr{\frac{1}{\sin [{\pi}(x+x'+iy)/2L]}}{\frac{1}{\sin [{\pi}(x-x'+iy)/2L]}}{\frac{1}{\sin [{\pi}(x-x'-iy)/2L}}{\frac{1}{\sin [{\pi}(x+x'-iy)/2L}}, \\
  \check{V}(x) &=& \matr{\sin\frac{\phi(L-x)}{2L}}{\cos\frac{{\phi}(L-x)}{2L}}{i\cos\frac{\phi x}{2L}}{i\sin\frac{\phi x}{2L}} \otimes \sigma_0.
\end{eqnarray}

Ths $s$-wave contribution of the regularized Green function can be found as
\begin{eqnarray}
 \check{G}_{\rm reg}^{ss}(x_0) &=&
  \lim_{\genfrac{}{}{0pt}{}{x'\rightarrow x_0}{y\rightarrow0}}[\check{G}_0(x_0,x';y)-g(x_0-x',y)] \nonumber \\ &=&
  \frac{i}{4\hbar vL}\check{V}(x_0)\matr{\frac{1}{\sin({\pi x_0}/{L})}}{-\sigma_x {\phi}/{\pi}}{\sigma_x {\phi}/{\pi}}{-\frac{1}{\sin({\pi x_0}/{L})}}\check{V}^{-1}(x_0)
  \label{eq:Gss}
\end{eqnarray}
Contributions from higher angular momentum modes are obtained by keeping higher terms of the Taylor series.
Therefore, we write the regularized Green function as
\begin{equation}
\label{eq:Taylor}
 \check{G}_{\rm reg}(\vec{r},\vec{r'})=\sum_{\mu,\nu}\frac{1}{\mu !}\frac{1}{\nu!} [(\vec{r}-\vec{r}_0)\overrightarrow{\nabla}]^{\mu}\check{G}_{\rm reg}(\vec{r}_0,\vec{r}_0) [\overleftarrow{\nabla}'(\vec{r}'-\vec{r}_0)]^{\nu}
\end{equation}
Here, $\overrightarrow{\nabla}$ acts to the right, on the first argument of the Green function, and $\overleftarrow{\nabla}'$ acts to the left, on the second argument of the Green function. We can simplify this expression by using the equations of motion of the regularized Green function, 
\begin{equation}
 -i\sigma\overrightarrow{\nabla}\check{G}_{\rm reg}=0,\qquad \check{G}_{\rm reg}(-i\overleftarrow{\nabla}'\sigma)=0
\end{equation}
from which immediately follows, that
\begin{equation}
 \partial_y \check{G}_{\rm reg}(\vec{r},\vec{r'}) = i\sigma_z \partial_x \check{G}_{\rm reg}(\vec{r},\vec{r'}), \qquad  \partial_{y'} \check{G}_{\rm reg}(\vec{r},\vec{r'}) =  \partial_{x'} \check{G}_{\rm reg}(\vec{r},\vec{r'})(-i\sigma_z).
\end{equation}
This enables us to rewrite Eq. \eqref{eq:Taylor} as
\begin{equation}
 \label{eq:Taylor2}
  \check{G}_{\rm reg}(\vec{r},\vec{r'})=\sum_{\mu,\nu} \tfrac{1}{\mu !}[(x-x_0)+i\sigma_z(y-y_0)]^{\mu}\check{G}_{\rm reg}^{\mu\nu} \tfrac{1}{\nu!}[(x'-x_0)-i\sigma_z(y'-y_0)]^{\nu},
\end{equation}
with
\begin{equation}
 \check{G}_{\rm reg}^{\mu\nu}=\partial_x^{\mu}\partial_{x'}^{\nu}\check{G}_{\rm reg}(\vec{r}_0,\vec{r}_0).
\end{equation}

We now give the explicit expressions for $ \check{G}_{\rm reg}^{\mu\nu}$ up to $d$-wave order. In order to keep the expressions short, we focus on the case, where the dot is placed in the middle of the sample $(x_0=L/2)$,
\begin{eqnarray*}
  \check{G}_{\rm reg}^{sp/ps} &=&\frac{i}{8\hbar v L^2}\check{V}(x_0)\matr{\pm\frac{\pi^2-3\phi^2}{6\pi}\sigma_x}{-\phi}{-\phi}{\pm\frac{\pi^2-3\phi^2}{6\pi}\sigma_x}\check{V}^{-1}(x_0), \\
 \check{G}_{\rm reg}^{pp} &=&\frac{i}{16\hbar v L^3}(\pi^2-\phi^2)\check{V}(x_0)\matr{1}{\frac{\phi}{3\pi}\sigma_x}{-\frac{\phi}{3\pi}\sigma_x}{-1}\check{V}^{-1}(x_0),\nonumber \\ 
 \check{G}_{\rm reg}^{sd/ds}&=&\frac{i}{16\hbar v L^3}(\pi^2-\phi^2)\check{V}(x_0)\matr{1}{-\frac{\phi}{3\pi}\sigma_x}{\frac{\phi}{3\pi}\sigma_x}{-1}\check{V}^{-1}(x_0), \\
 \check{G}_{\rm reg}^{pd/dp}&=&\frac{i}{32\hbar v L^4}\check{V}(x_0)\matr{\mp\frac{7\pi^4-30\pi^2\phi^2+15\phi^4}{60\pi}\sigma_x}{\phi(-3\pi^2+\phi^2)}{\phi(-3\pi^2+\phi^2)}{\mp\frac{7\pi^4-30\pi^2\phi^2+15\phi^4}{60\pi}\sigma_x}\check{V}^{-1}(x_0), \nonumber \\
 \check{G}_{\rm reg}^{dd}&=&\frac{i}{64\hbar v L^5}\check{V}(x_0)\matr{5\pi^4-6\pi^2\phi^2+\phi^4}{\phi\frac{-7\pi^4+10\pi^2\phi^2-3\phi^4}{15\pi}\sigma_x}{-\phi\frac{-7\pi^4+10\pi^2\phi^2-3\phi^4}{15\pi}\sigma_x}{-(5\pi^4-6\pi^2\phi^2+\phi^4)}\check{V}^{-1}(x_0).
\end{eqnarray*}
The element $\check{G}_{\rm reg}^{ss}$ is given by Eq.\ (\ref{eq:Gss}). Since $\check{V}(x_0)$ has no matrix structure in pseudospin space, it commutes with the $T$-matrix, and therefore does not play a role for the generating function.

Upon inserting \eqref{eq:Taylor2} into the generating function (Eq. \eqref{eq:F}), we find
\begin{equation}
 \delta \mathcal{F}=\ln \det [1-\mathbf{T} \check{\mathbf{G}}_{\rm reg}],
\end{equation}
where $\check{\mathbf{G}}_{\rm reg}$ is a (infinite) matrix with entries $G_{\rm reg}^{\mu\nu}$ ($\mu$,$\nu=0,1,...$) and the matrix $\mathbf{T}$ contains the elements $T^{\mu\nu}$
\begin{equation}
 T^{\mu\nu}=\int d^2\vec{r} d^2\vec{r'} \tfrac{1}{\mu!}[(x-x_0)-i\sigma_z(y-y_0)]^{\mu} T(\vec{r},\vec{r'}) \tfrac{1}{\nu!}[(x'-x_0)+i\sigma_z(y'-y_0)]^{\nu}.
  \label{eq:Tmunuapp}
\end{equation}
This is Eq.\ (\ref{eq:Tmunu}) of the main text.
In the limit that the size $R$ of the quantum dot is much smaller than $L$, this expansion is convergent, and it is a good approximation to limit the number of angular momentum ``channels'' involved in the expansion. In the simplest case, one takes into account the $s$-wave contribution $\mu=\nu=0$ only. This limit was considered in Ref.\ \onlinecite{kn:titov2010}. The accuracy can be improved by including contributions from higher angular momentum. The expressions derived above allow one to go up to the $d$-wave contribution.

In order to relate the object $T^{\mu\nu}$ of Eqs. \eqref{eq:Tmunu} or (\ref{eq:Tmunuapp}) to the $T$ matrix in the partial-wave expansion, it is instructive to compare the plane-wave and circular-wave basis sets of eigenstates of the free Dirac Hamiltonian $H_0=-i\mathbf{\sigma\nabla}$. The plane wave basis consists of eigenstates, which are labeled by their wavevector $\mathbf{k}=(k\cos\theta_k,k\sin\theta_k)$,
\begin{equation}
 \psi_{\vec{k}}(\vec{r})=\frac{e^{i\mathbf{kr}}}{\sqrt{2}}\vect{1}{e^{i\theta_k}}.
\end{equation}
Here and below, we restrict ourselves to positive energy solutions (conduction band) only. In the implementation, this can be achieved by choosing the gate potential of the dot to be negative. The circular-wave states are combined eigenstates of the Hamiltonian and the total angular momentum $L_z+\frac{1}{2}\sigma_z$, and are therefore labeled by their wavenumber $k$ and their half-integer angular momentum quantum number $m$,
\begin{equation}
 \psi_{km}(\vec{r})=\sqrt{\frac{k}{4\pi}}e^{im\theta}\vect{e^{-i\frac{\theta}{2}}J_{\left|m-\frac{1}{2}\right|}(kr)}{e^{i\frac{\theta}{2}}i\mathrm{sgn}(m)J_{\left|m+\frac{1}{2}\right|}(kr)}.
  \label{eq:psikm}
\end{equation}
Here, $J_{\nu}$ is the Bessel function of $\nu$-th order. The $2 \times 2$ matrix $\Psi_{k\mu}$ of the main text is related to the $\psi_{km}$ as $\Psi_{k\mu} = (\psi_{k,\mu+1/2},i \psi_{k,-\mu-1/2})$. The basis change between plane waves and circular waves is expressed through the equation
\begin{equation}
 \psi_{\vec{k}}(\vec{r})=\sqrt{\frac{2\pi}{k}}\sum_m i^{|m-\frac{1}{2}|} e^{-i(m-\frac{1}{2})\theta_k} \psi_{km}(\vec{r}).
\end{equation}

For the scattering problem in the infinite graphene sheet, we place the center of the quantum dot at the origin. We may express the $T$-matrix in the plane-wave basis through the $T$-matrix in real space as
\begin{eqnarray}
  \label{eq:Tkk1}
 T(\vec{k},\vec{k'}) =\frac{1}{2}\int d^2\mathbf{r} d^2\mathbf{r'}
  ( 1 \quad e^{-i\theta_k}) T(\mathbf{r},\mathbf{r'})
   \left(
    \begin{array}{c}
     1\\
     e^{i\theta_k'}
    \end{array}
   \right)
  \mathrm{exp}[ik(x'\cos\theta_k'+y'\sin\theta_k'-x\cos\theta_k-y\sin\theta_k)].
\end{eqnarray}
On the other hand, the same matrix elements can be also written in spherical wave basis,
\begin{eqnarray}
  \label{eq:Tkk2}
T(\vec{k},\vec{k'}) &=& \frac{2\pi}{k}\sum_{nm} (-i)^{|n-\frac{1}{2}|} i^{|m-\frac{1}{2}|} e^{i(n-\frac{1}{2})\theta_k} e^{-i(m-\frac{1}{2})\theta_k'}T_{nm}(k)
\end{eqnarray}
The partial-wave $T$ matrices $T_{mn}$ introduced here and $T_{\mu\nu}$ of the main text are related as
\begin{equation}
  T_{\mu\nu} = \left( \begin{array}{cc} T_{\mu+1/2,\nu+1/2} &
  i T_{\mu+1/2,-\nu-1/2} \\ -i T_{-\mu-1/2,\nu+1/2} &
   T_{-\mu-1/2,-\nu-1/2} \end{array} \right).
\end{equation}
Comparing Eqs.\ (\ref{eq:Tkk1}) and (\ref{eq:Tkk2}), we obtain a relation between the $T$-matrix in spherical wave basis and the $T$-matrix in real space. For example, for the $s$-wave channel we find
\begin{eqnarray}
 \frac{2\pi}{k}\matr{T_{\frac{1}{2},\frac{1}{2}}(k)}{iT_{\frac{1}{2},-\frac{1}{2}}(k)}{-iT_{-\frac{1}{2},\frac{1}{2}}(k)}{T_{-\frac{1}{2},-\frac{1}{2}}(k)}
  =\frac{1}{2}\int d^2\mathbf{r} d^2\mathbf{r'} T(\mathbf{r},\mathbf{r'})+\mathcal{O}(k),
\end{eqnarray}
so that
\begin{eqnarray}
 T^{00} &=&
  \lim_{k\rightarrow0}\frac{4\pi}{k}\matr{T_{\frac{1}{2},\frac{1}{2}}(k)}{iT_{\frac{1}{2},-\frac{1}{2}}(k)}{-iT_{-\frac{1}{2},\frac{1}{2}}(k)}{T_{-\frac{1}{2},-\frac{1}{2}}(k)} \nonumber \\ &=&
  \lim_{k\rightarrow0}\frac{4\pi}{k}
  T_{00}(k),
\end{eqnarray}
where, in the last line, we used the $2 \times 2$ matrix notation with integer indices used in the main text. The remaining identities in Eq.\ (\ref{eq:Tmunumunu}) follow similarly.

\section{Low-$k$-Limit of Scattering matrix}
\label{sec:Transfer}
In this appendix, we present the details how to determine the low-$k$ limit of the scattering matrix. As explained in the main text, we modify the scattering problem, such that we solve the scattering problem for particles with wavenumber $k_1=k+k'$ in the potential $U'=U+\hbar k' v$. We then need to relate the scattering matrix $S'$ for this problem, in which $U'$ is set to zero outside a certain cut-off radius $a$ to the scattering matrix $S$ for the original problem, in which $U$, not $U'$ is set to zero for $r > a$.

In order to relate these two scattering matrices, we need to calculate the scattering matrix of a potential step, in which the potential is zero for $r > a$ and equal to $u = \hbar k' v$ for $r < a$. Hereto, we expand the wavefunctions for $r < a$ and $r > a$ in terms of incoming and outgoing circular waves with wavenumber $k'+k$ and $k$, respectively,
\begin{eqnarray}
 \psi_{a-}(\vec{r}) &=&
  \sum_m (L_m^+ \psi_{k+k',m,+} (\vec{r})+ L_m^- \psi_{k+k',m,-}(\vec{r})), 
  \nonumber \\
 \psi_{a+}(\vec{r}) &=&
  \sum_m (R_m^+ \psi_{k,m,+} (\vec{r})+ R_m^- \psi_{k,m,-}(\vec{r})),
\end{eqnarray}
where the scattering states $\psi_{k,m,\pm}$ are obtained from the basis states $\psi_{km}$ in Eq.\ (\ref{eq:psikm}) by the replacement $J(kr) \to H^{\pm}(kr)$.
The coefficients $R_{m}^{\pm}$ and $L_{m}^{\pm}$ are related through the transfer matrix $\mathcal{T}$,
\begin{equation}
 \vect{R_m^{+}}{R_m^{-}}=\mathcal{T}_{m}(k,k+k',a)\vect{L_m^{-}}{L_m^{+}},
\end{equation}
which is easily calculated from continuity of the wave function at $r=a$,
\begin{equation}
 \mathcal{T}_{m}(k,k+k',a)= \frac{\pi a}{4i}\tilde{\mathcal{T}}_{m}(k,k+k',a)
  \sqrt{k(k+k')},
\end{equation}
where the $2\times2$ matrix $\tilde{\mathcal{T}}_{m}(k,k+k',a)$ has entries
\begin{eqnarray*}
 \tilde{\mathcal{T}}_{m}(k,k+k',a)_{11}&=&H^{(-)}_{|m|+\frac{1}{2}}(ka) H^{(+)}_{|m|-\frac{1}{2}}(ka+k'a)\nonumber- H^{(-)}_{|m|-\frac{1}{2}}(ka) H^{(+)}_{|m|+\frac{1}{2}}(ka+k'a), \\
 \tilde{\mathcal{T}}_{m}(k,k+k',a)_{12}&=&H^{(-)}_{|m|+\frac{1}{2}}(ka) H^{(-)}_{|m|-\frac{1}{2}}(ka+k'a)\nonumber-  H^{(-)}_{|m|-\frac{1}{2}}(ka) H^{(-)}_{|m|+\frac{1}{2}}(ka+k'a).
\end{eqnarray*}
The other entries are given by the relations
$\tilde{\mathcal{T}}_{m}(k,k+k',a)_{21}=-\tilde{\mathcal{T}}_{m}(k,k+k',a)_{12}^*$, $\tilde{\mathcal{T}}_{m}(k,k+k',a_0)_{22}= -\tilde{\mathcal{T}}_{m}(k,k+k',a)_{11}^*$. The transfer matrix is converted into reflection/transmission coefficients using the standard relations
\begin{eqnarray}
 \rho &=&\mathcal{T}_{12}(\mathcal{T}_{22})^{-1}, \nonumber \\
 \tau &=&(\mathcal{T}_{22})^{-1}, \nonumber\\
 \tau' &=&\mathcal{T}_{11}-\mathcal{T}_{12}(\mathcal{T}_{22})^{-1}\mathcal{T}_{21}, \nonumber\\
 \rho' &=&-(\mathcal{T}_{22})^{-1}\mathcal{T}_{21}. \nonumber\\
\end{eqnarray}
Upon taking the limit $k\rightarrow0$, we find
\begin{eqnarray}
 \rho_{m} &=&
  1+\frac{i\pi a^{2|m|}}{[(|m|-\frac{1}{2})!]^2 2^{2|m|-1}}\frac{H^{(-)}_{|m|+\frac{1}{2}}(k'a)}{H^{(-)}_{|m|-\frac{1}{2}}(k'a)}k^{2|m|} \nonumber 
 +\mathcal{O}(k^{2|m|+1}), \nonumber \\
 \tau_{m} &=& \tau'_{m}=  
  \frac{1}{(|m|-\frac{1}{2})!2^{|m|-\frac{3}{2}}}\frac{a^{|m|-\frac{1}{2}}}{\sqrt{k'}}\frac{1}{H^{(-)}_{|m|-\frac{1}{2}}(k'a)}k^{|m|}\nonumber +\mathcal{O}(k^{|m|+1}), \nonumber \\ 
  \rho'_{m} &=&-\frac{H^{(+)}_{|m|-\frac{1}{2}}(k'a)}{H^{(-)}_{|m|-\frac{1}{2}}(k'a)}+\mathcal{O}(k).
\end{eqnarray}
We can now relate the scattering matrices $S'$ for particles with energy $k'$ in the potential $U' = U+\hbar k' v$ and $U'=0$ for $r > a$ to the scattering matrix $S$ for particles with energy $k$ in the potential $U$, with $U = 0$ for $r > a$,
\begin{equation}
 S=\rho +\tau' ({1- S'\rho'})^{-1} S'\tau.
\end{equation}
Inserting the above expansions, we get an expression of the form
\begin{equation}
 S_{nm}(k)=\delta_{nm}+S^{(1)}_{nm}\, k^{|n|+|m|}+\mathcal{O}(k^{|n|+|m|+1}),
\end{equation}
coinciding with Eq. \eqref{eq:r-exp} in the main text. The coefficient $S^{(1)}$ depends on $a$, $k'$ and $S'$.
\end{widetext}

\end{document}